\documentclass[prl,reprint]{revtex4-1}

\usepackage{graphicx}
\usepackage{bm}
\usepackage{dsfont}
\usepackage{amsmath, amsthm, amssymb,mathrsfs}
\usepackage{setspace}
\usepackage[utf8]{inputenc}
\usepackage[toc,page]{appendix}
\usepackage{epstopdf}
\usepackage{hyperref}
\usepackage{mdframed}

\newcommand{\beq}{\begin{equation}}
\newcommand{\eeq}{\end{equation}}
\newcommand{\beqa}{\begin{eqnarray}}
\newcommand{\eeqa}{\end{eqnarray}}

\newcommand{\ket}[1]{\left| #1 \right\rangle}

\newcommand{\bra}[1]{\left\langle #1 \right|}

\newcommand{\av}[1]{\langle #1\rangle}

\newcommand{\e}{\mathrm{e}}

\newcommand{\mc}[1]{\mathcal{#1}}
\newcommand{\arFour}[4]{ \left( \begin{array}{cc} {#1} & {#2} \\ {#3} & {#4} \end{array} \right) }
\newcommand{\vFour}[4]{ \left( \begin{array}{c} {#1} \\ {#2} \\ {#3} \\ {#4} \end{array} \right) }

\DeclareMathOperator*{\argmin}{arg\,min}

\newcommand{\bqa}{\begin{eqnarray}}
\newcommand{\eqa}{\end{eqnarray}}

\definecolor{com}{rgb}{0.9,0.1,0.3}

\begin{document}
\title{\textbf{Structure in Multimode Squeezing:} \\
A Generalised Bloch-Messiah Reduction}
\author{Will McCutcheon}
\affiliation{Quantum Engineering Technology Laboratories - University of Bristol}
\date{\today}

\begin{abstract}
Methods to decompose nonlinear optical transformation vary from setting to setting, leading to apparent differences in the treatments used to model photon pair sources, compared to those used to model degenerate down-conversion processes. The Bloch-Messiah reduction of Gaussian processes to single-mode squeezers and passive (linear) unitaries appears juxtaposed against the practicalities of the Schmidt-decomposition for photon pair sources into two-mode squeezers and passive unitaries. Here, we present a general framework which unifies these forms as well as elucidating more general structure in multimode Gaussian transformations. The decomposition is achieved by introducing additional constraints into the Bloch-Messiah reduction used to diagonalise Gaussian processes, these constraints motivated by physical constraints following from the inequivalence of different physical degrees of freedom in a system, ie. the temporal-spectral degrees of freedom vs different spatial modes in a transformation. The result is the emergence of the two-mode squeezing picture from the reduction, as well as the potential to generalise these constraints to accommodate spectral imperfections in a source generating 3-mode continuous variable GHZ-like states. Furthermore, we consider the practical scenario in which a transformation aims to generate a multiphoton entangled state, whereby spatial modes provide desirable degrees of freedom, whilst undesired spectral mode structure contributes noise, and show that this spectral impurity can be efficiently modeled by finding an optimal low dimensional bases for its simulation. 
\end{abstract}
\maketitle


{\it Introduction.---} Nonlinear quantum optical processes provide the key resources for quantum metrology ~\cite{Pinel2013,Schnabel2017,Knott2016,Matthews2013,Luis2015,Demkowicz-Dobrzanski2015,Manceau2017,Steuernagel2004,StrategicMarketing2018}, photonic quantum computing~\cite{Raussendorf2003,Braunstein2004,Andersen2010,Kok2010,Bell2013}, and quantum communications~\cite{Bell2014b,McCutcheon2016}. Nonlinear processes at most quadratic/bilinear in the field operators admit a particularly concise representation since the dynamics can be fully represented by linear symplectic operations, allowing these \emph{Gaussian} operations to be modeled in a remarkably straightforward formalism~\cite{Balian1969,Dutta1995,Braunstein2005,Weedbrook2012,Adesso2014,Serafini2017a}, facilitating the modeling of parametric down conversion~\cite{Uren2006,Eckstein2011} and four wave mixing~\cite{Fang2013} in crystals, fibres~\cite{Alibart2006a,Clark2011,Smith2009}, waveguides~\cite{Yang2008,Brecht2011}, cavities~\cite{Vernon2015,Vernon2017} and photonic crystals~\cite{Sipe2004,Bhat2006}.
Gaussian quantum information thereby enables the assessment of key properties of states and processes such as photon pair purity~\cite{Grice1997,Grice,Law2000}, sensitivity in parameter estimation tasks~\cite{Jiang2014}, communication capacity~\cite{Vaidman1999}, and more general measures of quantum correlations~\cite{Gerke2015,Parker1999,Laurat2005,Adesso2016,Brandao2005,Cable2010}.

A necessary ingredient for using the tools of Gaussian quantum information to model nonlinear quantum optical processes is to choose, from the infinite-dimension spectral-temporal modes of an optical field undergoing a nonlinear process, an appropriate finite-dimensional basis upon which to use the tools of Gaussian quantum information. Luckily, the question of the existence of such a modal basis has been answered in the affirmative, with Bloch-Messiah (BM) reduction (also known as Euler decomposition), providing a natural canonical form~\cite{Braunstein2005,Wasilewski2006,Lvovsky,Ma1990} consisting of single-mode squeezers and passive unitaries. This Bloch-Messiah reduction establishes  strict limitations on the inconvertibility of Gaussian processes including the requirement that at least two single-mode squeezers (plus passive unitaries) are necessary to construct a two-mode squeezer~\cite{Braunstein2005}. And this equivalence of pairs of singlemode squeezers and two-mode squeezing has been well developed to model photon pair sources through the Schmidt decomposition~\cite{Law2000}. In this setting the book is widely believed to be closed, however a fairly strong assumption has been made regarding the degrees of freedom in the optical modes, i.e. each degree of freedom is treated \emph{equally}. This assumption is suitable on a single spatial mode, where only the spectral-temporal degrees of freedom are at play, or for a single spectral-temporal mode distributed over many spatial modes. However, where spatial and spectral degrees of freedom are both present, such as in the increasingly complex devices being developed experimentally, different degrees of freedom are, practically speaking, inequivalent. 

By introducing additional constraints into the Bloch-Messiah decomposition, we demonstrate a new family of canonical forms available for modeling Gaussian processes. We demonstrate that these provide a natural minimal modal basis for applying Gaussian quantum information techniques and recover known results for maximal squeezing in the presence of loss; in this framework we naturally recover the two mode squeezing picture as well as multimode squeezing pictures demonstrating that in some instances multi-mode squeezing is \emph{irreducible}; we then consider the generation of post-selected entangled states for photonic quantum information, presenting an efficient method to include parasitic spectral degrees of freedom in their simulation.

{\it Bloch-Messiah Reduction.---} An arbitrary Gaussian process in a single spatial mode is described by a general bogoliubov transformation,
\beq
\label{eqs:SingleModeBogoliubov}
\begin{split} 
 \hat b(\omega) &= \int d\omega'  C(\omega,\omega')\hat a(\omega') +S(\omega,\omega') \hat a^\dagger(\omega') \\
\end{split}
\eeq
where $\hat a(\omega)$ and $\hat b(\omega)$ ( $\hat a^\dagger(\omega)$ and $\hat b^\dagger(\omega)$) are annihilation (creation) operators for modes of frequency $\omega$, on the input and output spaces of the transformation.
Since the transformation must preserve the bosonic commutation relations ($[\hat b(\omega),\hat b^\dagger(\omega')]=\delta(\omega-\omega')$ and $[\hat b(\omega),\hat b(\omega')]=[\hat b^\dagger(\omega),\hat b^\dagger(\omega')]=0$), the integration kernel forms a linear symplectic operator on the operators, $\hat A (\omega) = \bigl( \hat a(\omega),\hat a^\dagger (\omega)\bigr)^T$ and $\hat B (\omega)= \bigl( \hat b(\omega),\hat b^\dagger (\omega)\bigr)^T$, and there exist bases of mode functions $\{\psi_n(\omega)\}_n$ ( $\{\phi_n(\omega)\}_n$ ) on the output (input) space, such that the integration kernels ($C(\omega,\omega')$ and $S(\omega,\omega')$) are simultaneously diagonalized,
\beq
\label{eqs:SingleModeEuler}
\begin{split} 
 \tilde b_n &= \sum_n \ C^D_{nn} \tilde a_n+ S^D_{nn} \tilde a_n^\dagger \,,
\end{split}
\eeq
where $\tilde b_n = \int d\omega \psi_n^*(\omega)\hat b(\omega)$ and $\tilde a_n = \int d\omega \phi_n^*(\omega)\hat a(\omega)$ are broadband mode operators, and $C^D$ and $S^D$ are real diagonal matrices. In this canonical basis the input mode $\tilde a_n$ is squeezed by $S^D_{nn}$ and output in mode $\tilde b_n$. 

From hereon we represent the integration kernels by matrix multiplication (though the continuous nature can be straightforwardly recovered), so that the mode operators form vectors, $\vec A = \bigl( \vec a,\vec a^\dagger \bigr)^T$ and in a slight abuse of notation we will use $\omega$ and $\omega'$ as indicies for their elements, $\hat a(\omega)\rightarrow \vec a_\omega$, so the elements of $C$ are $C_{\omega\omega'}$.
Singular value decomposition (SVD) along with symplecticity conditions allows (see Supplementary),
\beq
\label{eqs:SingleModeVectors}
\begin{split} 
 \vec B &=  \arFour{U}{0}{0}{U^*} \arFour{C^D}{S^D}{S^D}{C^D} \arFour{V}{0}{0}{V^*}^\dagger \vec A \\
\end{split}
\eeq

Introducing an addition spatial degree of freedom, indexed by $x$ and $x'$, the operator $\hat a_{\omega x}$ is the annihilation operator for a photon of frequency $\omega$ in spatial mode $x$, transforming as,
\beq
\label{eqs:TwoDOFVectors}
\begin{split} 
 \vec b_{\omega x} &= C_{\omega x \omega' x'} \vec a_{\omega' x'} + S_{\omega x \omega' x'} \vec a^\dagger_{\omega' x'} \,,
\end{split}
\eeq
with Einstein summation convention throughout. Conventional BM reduction amounts to flattening this tensorial structure, and results in the canonical form identical to Eqs.~\ref{eqs:SingleModeEuler} however, the basis modes, $\tilde b_n = \psi_{n \omega x} \vec b_{\omega x}$, (similarly for $\tilde a$) now have arbitrary structure on spatial and temporal degrees of freedom. Whilst this is suitable where one has arbitrary control over the two degrees of freedom simultaneously, in practice this is rarely, if ever, the case. In contrast, orthonormal bases for the temporal and spatial degrees of freedom independently would give rise to a more practical and insightful decomposition of the process at hand --- this is what we now present.

{\it Generalized Bloch-Messiah (GBM) Reduction.---}The form of the Bogloiubov transformation of Eqs.~\ref{eqs:TwoDOFVectors} shows the tensorial structure of the elements and motivates the message --- the indexes $\omega$ and $x$ should not be arbitrarily flattened. Fortunately, recent developments in tensor analysis provide a tool to achieve our aims --- \emph{Higher-order singular value decomposition} (HOSVD)~\cite{DeLathauwer2000}, (see Supplementary)
 --- a generalization of SVD to $n$-way tensors resulting in $n$ unitary matrices (reminiscent of $U$ and $V$) and an all-orthogonal core tensor, $S^\perp$ (likened to $S^D$). Decomposing the transformation kernel,
\beq
\label{eqs:TwoDOFCDecomp}
\begin{split} 
S_{\omega x \omega' x'} &= u^{(t)}_{\omega n} u^{(s)}_{x m} S^\perp_{n m n' m'} v^{(t)}_{\omega' n'} v^{(s)}_{x' m'} \,,
\end{split}
\eeq
where the $u^{(t)}_{\omega n}$,  $u^{(s)}_{x m}$, $v^{(t)}_{\omega' n'}$ and $ v^{(s)}_{x' m'}$, are the elements of unitary matrices $U^{(t)}$, $U^{(s)}$, $V^{(t)}$ and $V^{(s)}$, and the superscript $t$ ($s$) refers to the temporal (spatial) degrees of freedom. The core tensor is all-orthogonal, $S^\perp_{a_1 b c d}S^{\perp *}_{a_2 b c d} \propto \delta_{a_1 a_2}$, for each index, and ordered according to $||S^\perp_{1 b c d} S^{\perp *}_{1 b c d}||\geq||S^\perp_{2 b c d} S^{\perp *}_{2 b c d}||\geq||S^\perp_{3 b c d} S^{\perp *}_{3 b c d}||\geq...$, with $||\cdot||$ the Frobenius norm. 
This defines the GBM reduced form,
\beq
\label{eqs:TwoDOFcanonical}
\begin{split} 
\breve b_{n m} &= C^\perp_{n m n' m'} \breve a_{n' m'} + S^\perp_{n m n' m'} \breve a^\dagger_{n' m'} \,,
\end{split}
\eeq
where the basis modes, $ \breve b_{n m} = u^{*(t)}_{n\omega} u^{(s)*}_{m x} \vec b_{\omega x}$ and $ \breve a_{n m} = v^{(t)}_{n \omega} v^{(s)}_{m x} \vec a_{\omega x}$, now correspond to the $n$th temporal and $m$th spatial modes (see Fig~\ref{fig:GBM}), and $C^\perp$ is the transformation of $C$ by the same unitaries Eqs.~\ref{eqs:TwoDOFCDecomp}. Importantly, these modes are unitary on the temporal and spatial DOFs independently, $u^{(t)}_{n \omega} u^{*(t)}_{m \omega} = \delta_{nm}$ and similarly for modes in $U^{(s)}$, $V^{(t)}$ and $V^{(s)}$.

\begin{figure}[h!]
\label{fig:GBM}
\centering
\includegraphics[width=1\columnwidth]{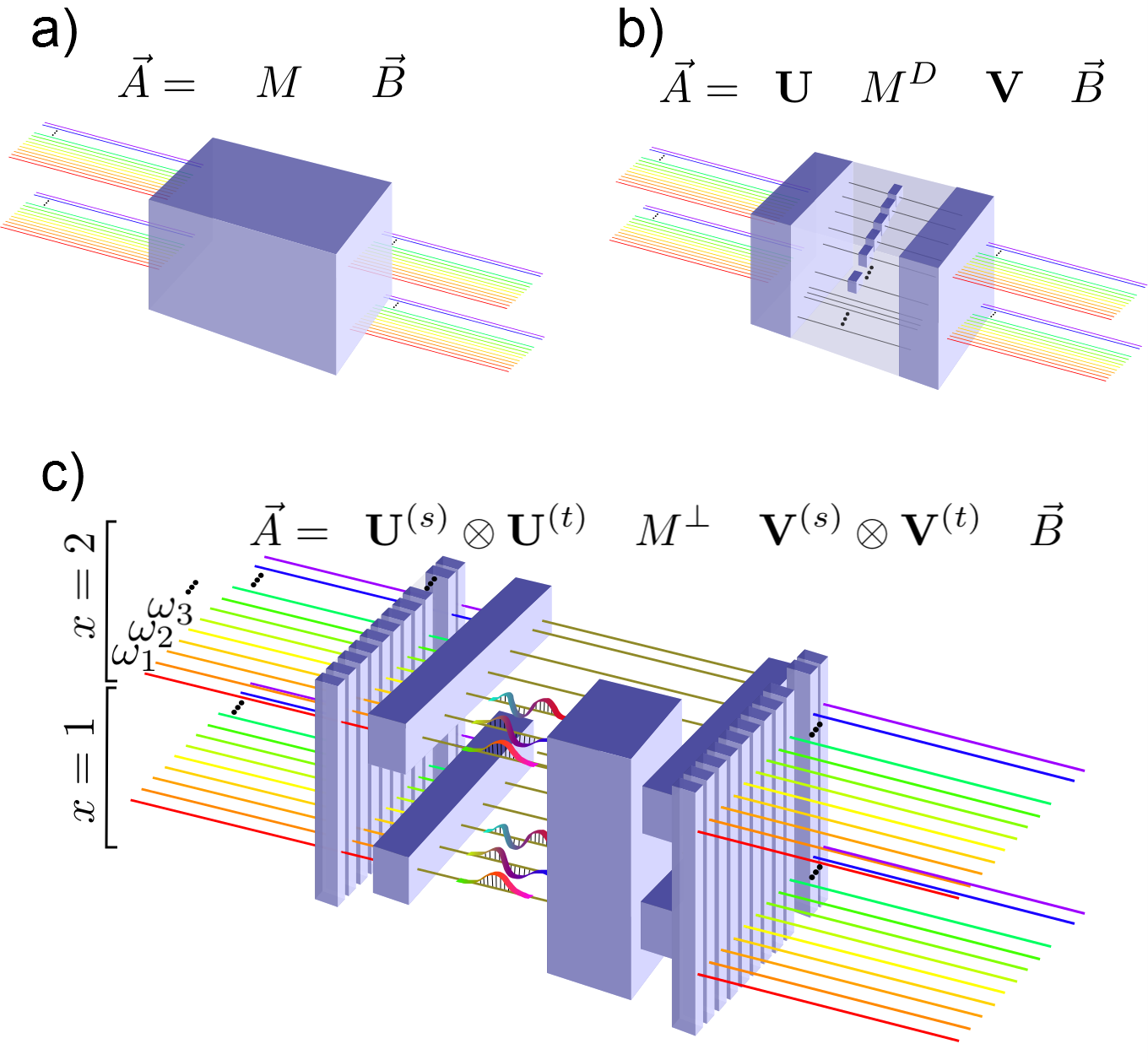}
\caption{
\label{fig:GBM}Decompositions of a Gaussian transformation $M$. a) The raw transformation, $M$ on continuous temporal DOFs and 2 spatial DOFs. b) The BM decomposition of $M$ into passive unitary transformations $\mathbf U$ and $\mathbf V$, and a diagonal active symplectic transformation $M^D$.
 c) A Generalised Bloch-Messiah decomposition of $M$ into the tensor product of unitaries $\mathbf U^{(s)}$ and $\mathbf U^{(t)}$ acting on the spatial and temporal degrees of freedom respectively, $M^\perp$, an all-orthogonal active symplectic transformation, and a further tensor product of spatial and temporal unitaries. Note that $M^\perp$ is active over only a finite subspace of the continuum of spectral modes. }
\end{figure}

{\it Single-mode Squeezing with loss.---}Consider a Gaussian transformation on two spatial modes initially in the vacuum state, where spatial mode one, $x=1$, is a bus mode we have access to, and spatial mode two, $x=2$, is an inaccessible loss mode. Transformation Eqs.~\ref{eqs:TwoDOFVectors} can be reduced to BM form Eqs.~\ref{eqs:SingleModeEuler} resulting in basis functions with support across both spatial modes. To evaluate the squeezing obtained in some spectral mode in the bus spatial mode we need an appropriate orthonormal basis over the spectral modes, however the component of the BM modes on the bus mode  $\{u_{n\omega 1}\}_n$ are not orthogonal. Consequently, we require to find an orthonormal basis by a Gram-Schmidt process. Once a suitable basis is chosen maximising the squeezing may be achieved (See supplementary). In contrast, we see GBM directly provides the natural minimal separable basis suitable for accessing spectral and spatial modes independently. In this GBM basis, one can recover the physical spatial mode basis by the inverse transformation $\mc U^{(s)\,-1}$ whilst still maintaining an orthogonal and ordered spectral mode basis. 
\footnote{We note that if one were to have access to both spatial modes, but constrained to only separable unitary operations, ie. an interferometer and a choosen local oscillator mode, despite the ordering of the core-tensor elements, the decomposition is over-constrained (requiring full unitarity) so the optimal choice for observing squeezing would not in general be expressed in the basis of our transformation. Instead, one would be interested in the Hitchcock’s rank decomposition, which is, in general computationally hard, and fails to provide the convinences of a unitary basis for the various degrees of freedom. Nonetheless, we believe pursuits of decompositions of this form would have potential merit for problems of this kind, though we do not pursue them here.}

{\it Two-mode Squeezing as an Irreducible Resource.---}
The two-mode squeezing formalism is intimately related to BM reduction, since two-mode squeezers can be reduced to a pair of single-mode squeezers and a passive unitary transformation~\cite{Braunstein2005}. The two-mode squeezing formalism comprises a method to reduce a system to two-mode squeezers, but not further, by \emph{choosing} to separate the transformation into disjoint integration regions. 
Where degenerate (singlemode) squeezing is vanishing, the energy matching constraints lead to a block antidiagonal structure in $S$, motivating this choice. The experimental availability of dichroic mirrors, acting as conditional SWAP operations between a pair of spatial modes and a bisection of the spectrum, encourages the decomposition to be addressed with this bisection in mind, thus separating the 'signal' and 'idler' halves of the spectrum. 
Here we see that GBM reduction introduces a constraint leading to the two-mode squeezing picture, demonstrating that two-mode squeezing is not only a possible decomposition, but \emph{irreducible} in this GBM picture. We perform the GBM reduction on the Hamiltonian level using a generalised Antoine-Takagi reduction, which in this particular setting is equivalent.

Consider the Antoine-Takagi decomposition of a two-mode squeezing Hamiltonian  (see Supplementary). The transformation generated by a Hamiltonian with vanishing linear dynamics (ie. interaction picture ) and neglecting time ordering~\cite{Lipfert2018}, takes the form
\beq
\label{eqs:AntoineTakGeneral}
\begin{split} 
\vec A &\rightarrow \exp\{i \arFour{0}{H}{-H^*}{0}\} \vec A \text{  ,} \,\,\,H=\arFour{0}{F_{JSA}}{F^T_{JSA}}{0} \, 
\end{split}
\eeq
with $F_{JSA}$ the joint-spectral-amplitude matrix.
Antoine-Takagi decomposition amounts to diagonalisation of $H$ by a congruence transformation $H= U H^D U^T$ (which exists since $H=H^T$) and results in a restricted class of GM reduced Gaussian transformations, general up to additional passive unitary transform, of the form,
\beq
\label{eqs:Antoine-BlochMessiah}
\begin{split} 
\vec A &\Rightarrow \arFour{U}{0}{0}{U^*} \arFour{\cosh H^D}{\sinh H^D}{\sinh H^D}{\cosh H^D}\arFour{U^\dagger}{0}{0}{U^T}  \vec A \\ 
\end{split}
\eeq
with $H^D$ a diagonal matrix imparting the squeezing parameters of the single-mode squeezers. 
Whilst this achieves BM reduction for arbitrary $F_{JSA}$, we contrast this to the two-mode squeezing picture which invokes the partial diagonalisation of $H$ via block diagonal unitaries,
\beq
\label{eqs:TwoModeAntoine}
\begin{split} 
H&
 = \arFour{U_s}{0}{0}{U_i}\arFour{0}{F^D}{F^D}{0}\arFour{U_s^T}{0}{0}{U_i^T}.
\end{split}
\eeq
This results in the transformation Kernel being of two-mode squeezing form,
\beq
\label{eqs:TwoModeAntoineKernel}
\begin{split} 
C&= \arFour{\cosh F^D}{0}{0}{\cosh F^D} \, \,,\, S= \arFour{0}{\sinh F^D}{\sinh F^D}{0} \, ,
\end{split}
\eeq
with $F^D$ diagonal. 
This can be reduced to BM form via a Hadamard transformation between the signal and idler halves of the spectrum. Constraining the reduction to prevent this Hadamard operation would see the two-mode squeezing picture prevail. To impose this constraint, we introduce an ancillary spatial mode, $x=2$, and mix these modes on a dichroic mirror, $D$, given by $D = \mathds{1}^{(s)}_{xx'}\otimes \mathds{1}^{+ (t)}_{\omega \omega'} + \hat X^{(s)}_{xx'}\otimes \mathds{1}^{-(t)}_{\omega \omega'}$, where $\mathds{1}^{+(t)}$ ($\mathds{1}^{-(t)}$) is a projector on the positive/signal (negative/idler) half of the spectrum, and $\hat X$ is the Pauli matrix, and we also apply this on the input modes to maintain the form of Eqs.\ref{eqs:AntoineTakGeneral}. GBM then leads to partial diagonlisation resulting in the two mode squeezing picture, Eqs.~\ref{eqs:TwoModeAntoineKernel}, via the unitary,
\beq
\label{eqs:dichroictwomode}
\begin{split} 
 \arFour{U_s}{0}{0}{U_i} \otimes \mathds{1}^{(s)} 
\end{split}
\eeq
which is separable with respect to the temporal and spatial degrees of freedom, unlike BM reduction which would require a non-separble unitary. The modes diagonalising the Hamiltonian in the BM picture, $\{ \tilde b_n \}_n$ consisting of symmetric, $\tilde b_{1} = \int d\omega (u^*_{sig}(\omega) +  u^*_{idl}(\omega))$, and anti-symmetric, $\tilde b_{2} = \int d\omega (u^*_{sig}(\omega) - u^*_{idl}(\omega))$, combinations of the signal and idler spectra, $u^*_{sig}(\omega)$ and $u^*_{idl}(\omega)$, yet the dichroic mirror forces these to become non-separable functions of spatial and temporal DOFs. In contrast, GBM reduction (including the dichroic mirror) results in the desired spectral modes $\tilde b_{sig(idl)} = \int d\omega u^*_{sig(idl)}(\omega) $ whilst the core tensor, Eqs.\ref{eqs:TwoModeAntoineKernel}, gains off-diagonal elements demonstrating the familiar two-mode squeezing picture. 

This method highlights two main distinctions. Firstly, with the inclusion of the dichroic mirror, the GBM reduction is constrained to result in the two-mode squeezing picture, establishing two-mode squeezing as an irreducible resource in this context. Secondly, whilst in this case the two-mode picture could be invoked by inspection, GBM reduction could be applied arbitrarily. In fact, one can apply the two-mode picture to a single mode squeezer by taking a Spontaneous-parametric down conversion source and introducing an appropriate dichroic mirror. One would find both single-mode and two mode squeezing present, which, in the limit of very narrow pump bandwidths reduces to form Eqs. \ref{eqs:AntoineTakGeneral} leaving just the two-mode squeezing contributions.  Furthermore, one can apply these methods to increasingly complex transformations, those having non-trivial conditional spectral operations generalising the dichroic mirror, and those extending over many spatial modes. We thus expect to see multi-mode squeezing in larger systems develop a rich structure, bounded only by the necessary structure of all-orthogonality of the core tensor. 

Finally, it is important to stress that whilst the core-tensors describing multi-mode squeezing can have a rich structure they are still significantly more practical to handle than the raw continuous mode transformations since they are both discrete, unlike the continuous spectral DOFs, and ordered so as one can truncate $S^\perp$, or $H^\perp$, to some small finite number of modes of interest that undergo squeezing.

{\it Three-mode Squeezing Picture for CV GHZ-like States.---}The continuous variable (CV) GHZ-like state can be generated by mixing three single-mode squeezed states on a \emph{tritter} (a three-mode Fourier transform)~\cite{VanLoock2001}. To achieve a minimum energy totally symmetric state the squeezing parameters and their quadratures must be choosen correctly.
We for now consider the states generated by an arbitrary Hamiltonian with the spectrum trisected into three regions forming the block-structure below, and introduce a three mode wavelength division multiplexer (generalisation of a dichoric mirror) to split these regions. Performing GBM on this Hamiltonian amounts to finding unitaries of the form $\text{diag}(U_1,U_2,U_3)$, resulting in the Hamiltonian,
\beq
\label{eqs:dichroictwomode}
\begin{split} 
\left( \begin{array}{ccc} {H_{11}} & {H_{12}} & {H_{13}}  \\ 
{H_{21}} & {H_{22}} & {H_{23}} \\
{H_{31}} & {H_{32}} & {H_{33}} 
\end{array} \right) \Rightarrow
\left( \begin{array}{ccc} {U_1 H_{11}U_1^T} & {U_1H_{12}U_2^T} & {U_1H_{13}U_3^T}  \\ 
{U_2H_{21}U_1^T} & {U_2H_{22}U_2^T} & {U_2H_{23}U_3^T} \\
{U_3H_{31}U_1^T} & {U_3H_{32}U_2^T} & {U_3H_{33}U_3^T} 
\end{array} \right) 
\end{split}
\eeq
For an ideal ensemble of disjoint GHZ-like states (analogous to the two-mode squeezing case) one must find that the diagonal terms are diagonalised by, $U_i H_{ii} U_i^T = H^D_{ii}\forall i$ , whilst these unitaries must also be the left and right singular vectors of the off diagonal terms, $U_i H_{ij} U_j^T = H_{ij}^D$ for $i\neq j$, and we require modewise symmetries $H_{ij}^D=H_{i'j'}^D \, \forall i\neq j$, $H_{ii}^D=H_{i'i'}^D \, \forall i$.
\footnote{For 3-mode GHZ-like state with vanishing local squeezing one requires that  $H^D_{ii} = H_{ij}^D -1/4\text{Ln}[\e^{6 H_{ij}^D}(2+\e^{6 })/(1+2\e^{6H_{ij}^D})]$ in the element-wise sense. Similiar expressions can be derived for GHZ states of arbitrary dimension.}. General Hamiltonians will however suffer spectral imperfections leading to a GBM form having off-diagonal terms in the submatrices of the core-tensor leading to parasitic single-mode and two-mode squeezing contributions. This GBM form allows these imperfections to be highlighted and modeled in a low dimensional setting. 
%

{\it Multi-mode Squeezing For Generating Large Photonic Entangled States.---}
When generating photonic entangled states, structured degrees of freedom in a system are integral for defining subsystems between which entanglement may exist. 
 For example we may aim to find a low dimensional basis for the undesirable temporal degrees of freedom arising from imperfect sources whilst maintaining the spatial and/or polarisation degrees of freedom in their physical basis.

For a general transformation conventional Bloch-Messiah gives us a practical means to expand the state into a Schodinger picture state in Fock space, and grouping terms by the total photon number we have,
\beq
\label{eqs:TwotwomodeSchro}
\begin{split} 
\ket{\psi_{BM}}&= \prod_d \tilde S_d(s_d)\ket{vac}
=\sum_N^{N_T} \ket{\psi^{(N)}_{BM}}\, ,
\end{split}
\eeq
where $\tilde S_d(s_d)$ are squeezing operators acting on the modes $\{\tilde b_n\}_n$ in the BM basis defined by the unitary $U$ such that $S=US^D V^T$. When post-selecting a state with fixed photon number, using a projector diagonal in the photon number basis, we need only consider the block arising from the corresponding term in the outer summation, ie. some fixed $N$.
Then using GBM reduction of the transformation to find the truncated unitary $\tilde U^{(t)}$, which transforms to a minimal spectral basis, we may construct $\bold U = (\tilde U^{(t)}\otimes \mathds{1}^{(s)} \otimes \mathds{1}^{(p)}) U^\dagger$. Then for some postselection projector $\hat P$ we can transform our Schrodinger picture state to give a postselected state in a physical basis 
\beq
\label{eqs:SchroEPR}
\begin{split} 
\ket{\psi_{phys}^{(N)}}= \hat P \bold{U}_N \ket{\psi^{(N)}_{BM}}
\end{split}
\eeq
where $\bold U_N$ is the irrep of $\bold U$ acting on the $N$ photon fock space. Expanding the initial state over the fock basis $\ket{\psi^{(N)}_{BM}}= \sum_{\vec m} \alpha_{\vec m} \ket{\vec m}$ and associating each postselected state fock state to an element of our abstract Hilbert space (eg. computation qubit basis)  $\ket{\vec n} \cong \ket{\psi(\vec n)}$, we have, 
\beq
\label{eqs:SchroEPRFock}
\begin{split} 
\ket{\psi_{comp}^{(N)}}= \sum_{\vec m \in \mc I, \vec n \in \mc P} \alpha_{\vec m}\ket{\psi(\vec n)}  \bra{\vec n} \bold{U}_N \ket{\vec m}
\end{split}
\eeq

 In larger systems, constructing the irrep $\bold{U}_N$ becomes impractical but efficient ways to characterize these transition elements, $\bra{\vec n} \bold{U}_N \ket{\vec m}$, are available~\cite{Arkhipov2014} and may be expressed via permanents. Whilst full characterisation of these elements is known to be hard, when both $|\mc I |$ and $|\mc P|$ are constant calculating Eqs.~\ref{eqs:SchroEPRFock} is only polynomial in the number of modes. In particular, choosing the minimal number of temporal modes, those rows of $U^{(t)}$, allows this expansion to be done \emph{most} efficiently.

{\it Fidelity of Truncated States.---}When considering some finite dimensional truncated approximation of the ideal state, the natural question to ask is what fidelity to my true infinite dimensional state do I achieve? In particular, is this basis of $U^{(t)}$ the optimal basis in which to truncate my system? 

Considering the state obtained by acting zero-mean Gaussian operation (eg. Eqs.\ref{eqs:TwoDOFVectors}) on the vacuum, we can evaluate the total photon number operator $\hat H_{tot} = \sum_i \vec{a}^\dagger_i \vec{a}_i$ to find,
\beq
\label{eqs:TruncaterFid}
\begin{split} 
\av{\hat H_{tot}} = \sum_{nm} |S_{nm}|^2 \, \approx \sum_{nm} |\tilde S_{nm}|^2 \, ,
\end{split}
\eeq
with $S$ and $\tilde S$ in an arbitrary, possible separable, basis, eg. $n\sim x\omega$, and $\tilde S$ is a truncation of $ S$. Of course this is the squared Frobenius norm, and simulations truncated to some subspace to maximize this, the GBM reduction in particular, will capture the maximum number of photons for any subspace of this dimension. Similarly, a biphoton state obtained by expanding the propagator to first order, results in a fidelity of the full non-normalised biphoton state with respect to that generated by a truncated Hamiltonian, of,
\beq
\label{eqs:TruncaterFid}
\begin{split} 
\av{\tilde \psi_{bi}| \psi_{bi}} = 2 \sum_{nm} |\tilde H_{nm}|^2 \, .
\end{split}
\eeq
which is maximised by truncating a Hamiltonian using the generalised Antoine-Takagi reduction method. 

{\it Conclusion.---}We demonstrated that the common two-mode squeezing picture results from adding further constraints to the methods of Bloch-Messiah reduction. These constraints impose separability between different distinct degrees of freedom, which corresponds to physically motivated practicalities in experimental settings. The resulting bases obtained allow for greater ease dealing with systems in which only certain spatial modes are accessible. Where temporal degrees of freedom constitute experimental noise, truncating the system in a GBM basis facilitates efficient low-dimensional simulation. GBM bases, and generalised Antoine-Takagi bases, are seen to optimise different figures of merit, the total photon number in the truncated subspace, and the biphoton state fidelity, respectively. 

\onecolumngrid
\newpage

\begin{appendices}

\section{Conventional Bloch-Messiah Reduction}
\label{app:Bloch-Messiah}

\subsubsection{Continuous Bloch-Messiah Reduction on a Single Spatial Mode }
In the absence of spatial modes, a Gaussian mode transformation takes the form

\beq
\label{eqs:SingleModeTransition}
\begin{split} 
\left( \begin{array}{c}
\hat b(\omega)\\
 \hat b^\dagger(\omega) \end{array} \right) &= \int d\omega'
\left( \begin{array}{cc}
C(\omega,\omega')  & S(\omega,\omega')\\
S^*(\omega,\omega') & C^*(\omega,\omega') \end{array} \right) 
\left( \begin{array}{c}
\hat a(\omega')\\
 \hat a^\dagger(\omega') \end{array} \right)  \\
\text{so that} \quad \hat b(\omega) &= \int d\omega'  C(\omega,\omega')\hat a(\omega') +S(\omega,\omega') \hat a^\dagger(\omega') \,.
\end{split}
\eeq
The Bloch-Messiah reduction demonstrates the simultaneous diagonalisation of $C$ and $S$ by considering the constraints on the system due to preservation of the commutators,

\beq
\label{eqs:SingleModeComutatorConstraint1}
\begin{split} 
\delta(\omega-\omega') &= [\hat b(\omega), \hat b^\dagger(\omega')] \\
&= \int d\omega''    C(\omega,\omega'') C^*(\omega',\omega'') - S(\omega,\omega'')S^*(\omega',\omega'') \\
\end{split}
\eeq
\beq
\label{eqs:SingleModeComutatorConstraint2}
\begin{split} 
0&= [\hat b(\omega), \hat b(\omega')] \\
&= \int d\omega''    C(\omega,\omega'')S(\omega',\omega'') - S(\omega,\omega'')C(\omega',\omega'') \\
\end{split}
\eeq
Decomposition of $C(\omega,\omega')$ to the form,
\beq
\label{eqs:CDecompContinuous}
\begin{split} 
C(\omega,\omega') =\sum_n \lambda_n \psi_n(\omega) \phi_n(\omega')\, ,
\end{split}
\eeq
with the orthonormal basis functions $\{ \psi_n(\omega)\}_n$ and $\{ \phi_n(\omega)\}_n$ defining unitary basis transformations, and can be found by solving the integro-eigenvlaue problems,
\beq
\label{eqs:CIntegroEigContinuous}
\begin{split} 
\int d\omega'' d\omega' C(\omega,\omega'')C^*(\omega',\omega'') \psi_n(\omega') =\lambda_n \psi_n(\omega)\\
\int d\omega'' d\omega' C^*(\omega'',\omega)C(\omega'',\omega') \phi_n(\omega') =\lambda_n \phi_n(\omega)\, .
\end{split}
\eeq
The discrete broadband modal bases, $\{ \hat b_n\}_n$, and $\{ \hat a_n\}_n$, on the input and output spaces respectively, defined by the envelopes, $\phi_n$ and $\psi_n$,
\beq
\label{eqs:ModeEnvelopes}
\begin{split} 
\hat a_n = \int d\omega \phi_n(\omega)  a(\omega) \quad \text{and} \quad \hat b_n = \int d\omega \psi_n(\omega) b(\omega)\, ,
\end{split}
\eeq
, also the inverse relations,
\beq
\label{eqs:InverseModeEnvelopes}
\begin{split} 
a(\omega) = \sum_n  \phi^*_n(\omega)  \hat a_n\quad \text{and} \quad b(\omega) = \sum_n  \psi^*_n(\omega)  \hat b_n\, .
\end{split}
\eeq
form bases which codiagonalises $C(\omega,\omega')$ and $S(\omega,\omega')$. Condition Eqs.~\ref{eqs:SingleModeComutatorConstraint1} implies that the $\psi_n$ diagonalise the intergration kernel $\int d\omega''    S(\omega,\omega'')S^*(\omega',\omega'')$, whilst the commutator $ [\hat b^\dagger(\omega), \hat b(\omega')]$ leads to diagonalisation of $\int d\omega'' S(\omega'',\omega)S^*(\omega'',\omega') $ by $\phi_n^*$. This continuous form of singular value decomposition can be more practically notated using discrete matrices which we next present.

\subsubsection{Discretising Time: Matrix Form}
We will denote the continuous linear transformations, by discretised sums
\beq
\label{eqs:MatrixBogoliubov}
\begin{split} 
\hat b(\omega) &= \int d\omega'  C(\omega,\omega')\hat a(\omega') +S(\omega,\omega') \hat a^\dagger(\omega') \qquad   \cong \qquad \vec b_\omega =  \sum_{\omega'} C_{\omega \omega'} \vec a_{\omega'}+S_{\omega \omega'} \vec a^\dagger_{\omega'} \, ,
\end{split}
\eeq
over the vectors of mode operators as $\vec a = (\hat a(\omega_1),\hat a(\omega_2),...,\hat a(\omega_d))^T$, and $\vec b = (\hat b(\omega_1),\hat b(\omega_2),...,\hat b(\omega_d))^T$. We shall from now on adopt einstein summation convention. The vectors of envelop functions $ \Phi (\omega) = (\phi_1(\omega),\phi_2(\omega),...\phi_d(\omega) )^T$,  $\Psi (\omega) = (\psi_1(\omega),\psi_2(\omega),...\psi_d(\omega) )^T$ become unitary \emph{matrices} with $n$ and $\omega$ indexing the elements, $ \Phi_{n\omega}$ and $\Psi_{n\omega}$. Consequently, we can write the mode transformations more simply as,
\beq
\label{eqs:ModeEnvelopesMatrix}
\begin{split} 
\tilde a_n =  \Phi_{n\omega} \cdot \vec a_\omega \\
\tilde b_n = \Psi_{n\omega} \cdot \vec b_\omega\\
\end{split}
\eeq
and
\beq
\label{eqs:InverseModeEnvelopes}
\begin{split} 
\vec a_\omega= \Phi^*_{n\omega}  \cdot \tilde a_n\\
\vec b_\omega = \Psi^*_{n\omega}  \cdot \tilde b_n\\
\end{split}
\eeq
where repeated indices imply summation. Completeness and unitarity ensure that $\Psi^\dagger \Psi=\Psi \Psi^\dagger=\Phi^\dagger \Phi= \Phi\Phi^\dagger= \mathds{1}$.
With this notation we can write the discritised Bogliubov transformation as simply,
\beq
\label{eqs:SingleModeTransitionMatrix}
\begin{split} 
\vec B =\left( \begin{array}{c}
\vec b\\
 \vec b^{(\dagger)} \end{array} \right) &= 
\left( \begin{array}{cc}
C  & S\\
S^* & C^* \end{array} \right) 
\left( \begin{array}{c}
\vec a\\
 \vec a^{(\dagger)} \end{array} \right)  = M \vec A \, ,
\end{split}
\eeq
where,
\beq
\label{eqs:SymplecticMats}
\begin{split} 
\vec b^{(\dagger)} &= (\hat b_1^\dagger, \hat b_2^\dagger ...)^T \quad , \quad \vec a^{(\dagger)} = (\hat a_1^\dagger, \hat a_2^\dagger ...)^T\\
\vec B&= \left( \begin{array}{c}
\vec b\\
 \vec b^{(\dagger)} \end{array} \right)  ,  
 \quad \vec A=\left( \begin{array}{c}
\vec a\\
 \vec a^{(\dagger)} \end{array} \right)  \\
 M &= \left( \begin{array}{cc}
C  & S\\
S^* & C^* \end{array} \right) 
,
\end{split}
\eeq
and the commutator conditions can be written
\beq
\label{eqs:SingleModeComutatorConstraintMatrix}
\begin{split} 
\delta_{\omega \omega'}&=   C_{\omega \omega''} C^*_{\omega'\omega''} - S_{\omega\omega''}S^*_{\omega'\omega''} \Rightarrow CC^\dagger-SS^\dagger = \mathds{1}\\
0 &= C_{\omega\omega''} S_{\omega'\omega''} - S_{\omega\omega''}C_{\omega'\omega''} \Rightarrow CS^T-SC^T = 0 \, .
\end{split}
\eeq
Equivalently, we can write the commutation relations as,
\beq
\label{eqs:SymplecticCommutation}
\begin{split} 
[\vec A_i,\vec A_j] &= \Omega_{ij}\\
\text{where} \quad \Omega&= \left( \begin{array}{cc}
0  & \mathds{1}\\
-\mathds{1} & 0\end{array} \right) \, .
\end{split}
\eeq

This allows quick confirmation of the symplectic structure of linear cannonical transformations $M$ since,
\beq
\label{eqs:SymplecticCommutationCondition}
\begin{split} 
M_{ij} \Omega_{jk} (M^T)_{kl}&=M_{ij}[\tilde A_j,\tilde A_k](M^T)_{kl} \\
&= M_{ij} \tilde A_j  \tilde A_k M_{lk} -  M_{ij}\tilde A_k \tilde A_j M_{lk}\\
&= (M \tilde A)_i  (M \tilde A)_l -  (M\tilde A)_l (M \tilde A)_i\\
&= [\tilde B_i,\tilde B_l] = \Omega_{il} \\
\Rightarrow  M\Omega M^T&=\Omega \,.
\end{split}
\eeq
Furthemore, this property of symplectic matrices leads immediately to the definition of the inverse since $\Omega^2 = -\mathds{1}$, we have,
\beq
\label{eqs:SymplecticInverse}
\begin{split} 
  M \Omega M^T \Omega &=\Omega^2 =-\mathds{1}\\
\Rightarrow M^{-1} &= -\Omega M^T \Omega \, ,
\end{split}
\eeq
so that,
\beq
\label{eqs:SingleModeTransitionMatrixAdjoint2}
\begin{split} 
M^{-1}=\left( \begin{array}{cc}
C  & S\\
S^* & C^* \end{array} \right) ^{-1}
&=-\Omega \left( \begin{array}{cc}
C  & S\\
S^* & C^* \end{array} \right)^T \Omega \\
  &= 
\left( \begin{array}{cc}
C^\dagger  & -S^T\\
-S^\dagger & C^T \end{array} \right)\\
\end{split}
\eeq
Confirming that the commutators are preserved under the inverse map, amounts to confirming the symplectic structure of $M^{-1}$, by evaluating $M^{-1}\Omega (M^{-1})^T=\Omega$. This adds two further constraints to the system so in total we have,
\begin{align}
 CC^\dagger-SS^\dagger &= \mathds{1} \label{eqs:AllConstraints1}\\
 CS^T-SC^T &= 0 \label{eqs:AllConstraints2} \\
 C^\dagger C-S^T S^* &= \mathds{1} \label{eqs:AllConstraints3} \\
 C^\dagger S-S^TC^* &= 0 \label{eqs:AllConstraints4} \, .
\end{align}
\subsubsection{Joint SVD}
From constraint \ref{eqs:AllConstraints1} we see that a unitary $U$ exists such that $U^\dagger CC^\dagger U = C_D^2$ is diagonal whilst $U^\dagger S S^\dagger U=S_D^2$ must also be diagonal, with $C_D^2-S_D^2 = \mathds{1}$. From \ref{eqs:AllConstraints3} we see that there exists a $V$ such that $V^\dagger C^\dagger C V = C_D^2$ and $V^\dagger S^T S^* V = S_D^2$. We may therefore have that $C=U C_D V^\dagger$ and $S=US_DV^T$.

These unitaries diagonalising $C$ and $S$ are going to be exactly the unitary transformation to broadband modes that we will use.  Explicitly, taking $\Phi_{nt} = (V^\dagger)_{nt}$ and $\Psi_{nt}=(U^\dagger)_{nt}$ we have,

\beq
\label{eqs:SingleModeBlochMessiahReduction}
\begin{split} 
\left( \begin{array}{c}
\vec b\\
 \vec b^{(\dagger)} \end{array} \right) &= 
\left( \begin{array}{cc}
U  & 0\\
0 & U^* \end{array} \right) 
\left( \begin{array}{cc}
C_D  & S_D\\
S_D^* & C_D^* \end{array} \right)
\left( \begin{array}{cc}
V  & 0\\
0 & V^* \end{array} \right)^\dagger  
\left( \begin{array}{c}
\vec a\\
 \vec a^{(\dagger)} \end{array} \right)  \\
\Rightarrow \left( \begin{array}{c}
U^\dagger \vec b\\
 U^T \vec b^{(\dagger)} \end{array} \right) &= 
\left( \begin{array}{cc}
C_D  & S_D\\
S_D^* & C_D^* \end{array} \right)
\left( \begin{array}{c}
V^\dagger \vec a\\
V^T \vec a^{(\dagger)} \end{array} \right) \\
= \left( \begin{array}{c}
\tilde b\\
 \tilde b^{(\dagger)} \end{array} \right) &=
\left( \begin{array}{cc}
C_D  & S_D\\
S_D^* & C_D^* \end{array} \right)
\left( \begin{array}{c}
\tilde a\\
\tilde a^{(\dagger)} \end{array} \right)  \\
\end{split}
\eeq

\subsubsection{Takagi Factorization}
Above, we proved the existence of such a decomposition, however, singular value decomposition is not unique and can lead to choices of $V$ which do not transform $C$ (or $S$) into real diagonal matrices. For instance, if the $i$th singular value $c_i$ of $C_D$ has multiplicity $n_i$, then we consider any block diagonal unitary matrix, $O$, with the same block structure as $C_D$ (also $S_D$).
\beq
\label{eqs:BlockSVDUnitary}
\begin{split} 
C_D &= \bigoplus_i c_i \mathds{1}_{n_i}\\
O&= \bigoplus_i O_i \qquad \dim O_i = n_i \\
\end{split}
\eeq
Therefore, $O C_DO^\dagger=O^\dagger C_D O =C_D$, and the singular value decomposition is non-unique, with an orbit of valid solutions generated by $O$.
\beq
\label{eqs:BlockSVDOrbit}
\begin{split} 
C&=U C_D V^\dagger\\
&= U O O^\dagger C_D O O^\dagger V^\dagger\\
&= (UO) C_D (VO)^\dagger
\end{split}
\eeq
We aim to use Takagi factorisation to find SVDs of $C$ and $S$ using the same unitary operators $U$ and $V$ that achieve
\beq
\label{eqs:SVDCondition}
\begin{split} 
C= U C_D V^\dagger \qquad \text{and}\qquad S= US_DV^T\,.
\end{split}
\eeq
We instead start with a general SVD of $C$, giving $U$ and $V_C$, then find the SVD of $S$ corresponding to this $U_C$,
\beq
\label{eqs:generalSVDC}
\begin{split} 
C= U_C C_D V_C^\dagger \qquad \text{and}\qquad S= U_CS_DV_S^\dagger\,.
\end{split}
\eeq
 We remind that with degenerate SVDs it is not in general true that $V_C=V_S^*$. Condition~\ref{eqs:AllConstraints3} requires that the operator $G=V_C^\dagger V_S^*$ commutes with $C_D^2$ and $S_D^2$ (and thus also  $C_D$ and $S_D$) by,
\beq
\label{eqs:GCommutes}
\begin{split} 
  \mathds{1} &= C^\dagger C-S^T S^* \\
&= V_C C_D^2 V_C^\dagger - V_S^* S_D^2  V_S^T\\
&= V_C (\mathds{1}+S_D^2) V_C^\dagger - V_S^* S_D^2  V_S^T\\
\therefore  \quad  S_D^2 V_C^\dagger V_S^* &= V_C^\dagger V_S^* S_D^2 \qquad  \Rightarrow [S_D, G]=0 ,
\end{split}
\eeq
which guarantees it has the block-diagonal form of $O$, Eqs.~\ref{eqs:BlockSVDUnitary}. From condition~\ref{eqs:AllConstraints4} we also have that $G=G^T$ is symmetric (on the support of $S_D$ as we will see below), by,
\beq
\begin{split}
\label{eqs:GSymmetric}
0&= C^\dagger S-S^TC^* \\
&= V_C C_D S_D V_S^\dagger - V_S^* S_D C_D  V_C^T\\
&= C_D S_D V_S^\dagger V_C^* - V_C^\dagger V_S^* S_D C_D \\
&= C_D S_D ( G^T - G) \\
\end{split}
\eeq
and since it is unitary, it's Takagi decomposition takes the form $G=D \mathds{1} D^T = DD^T$ with $D$ also being of block diagonal form Eqs.~\ref{eqs:BlockSVDUnitary}. Consequently, by the structure of $G$ and thus $D$, any unitaries $UD$ and $VD$ give valid SVDs of $C$ and $S$ in accordance with Eqs.~\ref{eqs:BlockSVDOrbit}. Taking $U_C D$ and $V_C D$ for the unitary diagonalising $C$, and $U_C D$ and $V_S D$ for that of $S$ we have $V_C D = (V_S  D)^*$ so Eqs.~\ref{eqs:SVDCondition} holds for $U = U_CD$ and $V=V_C D$.

\subsubsection{Singular $S$ and Truncated Spaces}
Often, not all the modes are squeezed, and thus $S_D$ has some singular values which are zero (equivalently, $C_D$ has some singular values which are one). In this case a more compact expression of the SVD of $S$ can be achieved using just the rows of $U$ and $V_S$ which have support on the nonzero elements of $S_D$,
\begin{align}
\label{eqs:TruncatedSVD}
S= \tilde U \tilde S_D \tilde V_S^\dagger
\end{align}
where $\tilde S_D$ is the square diagonal matrix containing only the $d$ nonzero singular values of $S$ and the tildes on $\tilde U$ and $\tilde V_S$ indicate the rectangular matrix containing just the first $d$ rows. Returning to the Takagi factorisation in this case allows us to write 
\beq
\label{eqs:GCommutesTruncated}
\begin{split} 
  \mathds{1} &= C^\dagger C-S^T S^* \\
&= V_C C_D^2 V_C^\dagger - \tilde V_S^* \tilde S_D^2  \tilde V_S^T\\
&= V_C (\mathds{1} + \tilde S_D^2\oplus \emptyset) V_C^\dagger - \tilde V_S^* \tilde S_D^2  \tilde V_S^T\\
0 &= \tilde V_C \tilde S_D^2 \tilde V_C^\dagger -  \tilde V_S^*  \tilde S_D^2  \tilde  V_S^T\\
\therefore  \quad \tilde S_D^2 \tilde V_C^\dagger \tilde V_S^* &=\tilde  V_C^\dagger \tilde V_S^* \tilde S_D^2\\
 \Rightarrow 0&=[\tilde S_D,\tilde  G] ,
\end{split}
\eeq
with $\tilde G = \tilde  V_C^\dagger \tilde V_S^*$, so we need only consider the rectangular sub-matrices $\tilde  V_C$ and $ \tilde V_S$. Similarly, the symmetry of this reduced $\tilde G$ follows,
\beq
\begin{split}
\label{eqs:GSymmetricTruncated}
0&= C^\dagger S-S^TC^* \\
&= V_C C_D U^\dagger \tilde U \tilde S_D \tilde V_S^\dagger - \tilde V_S^* \tilde S_D \tilde U^T U^* C_D  V_C^T\\
&=\tilde V_C \tilde C_D \tilde S_D \tilde V_S^\dagger - \tilde V_S^* \tilde S_D \tilde C_D \tilde V_C^T\\
&=\tilde C_D \tilde S_D \tilde V_S^\dagger \tilde V_C^* - \tilde V_C^\dagger \tilde V_S^* \tilde S_D \tilde C_D \\
&= \tilde C_D \tilde S_D ( \tilde G^T - \tilde G) \\
\end{split}
\eeq
Thus use of $\tilde G$ resolves the issue that $G$ would otherwise be symmetric only on the support of $S_D$.

\subsubsection{Continuous Time Methods}
In general, numeric solutions to the above method applied to functions over continuous time will have non vanishing singular values in $S_D$ and an appropriate cutoff must be choosen, beyond which the remaining functions spanning the space experience only unitary behaviour. For instance, a matrix $C$ displaying some unitary behaviour (having some unit singular values), can be expressed as,
\beq
\label{eqs:UnitaryCPart}
\begin{split} 
C&= \tilde U \tilde C_D \tilde V^\dagger + \tilde U^\perp (\tilde V^\perp)^\dagger\\
\text{where} \qquad U &= (\tilde U, \tilde U^\perp) \quad \text{and}\quad V= (\tilde V,\tilde V^\perp)
\end{split}
\eeq
The \emph{squeezing modes} can be expressed via only considering the elements $\tilde U$, $\tilde C_D$ and $\tilde V$. 

In general, it is often more practical to first consider the action of the squeezed modes via decomposition of $S$, before inferring the remaining unitary behaviour in $C$. What's more, this can overcome some challenges associated with evaluating the full the transformation, since the unitary behaviour can be efficiently computed after evaluating the behaviour of the transformation on the finite squeezing modes.

\section{Bloch-Messiah Algorithm: Single Spatial Mode}
\label{app:BMAlgorithm}
We summarise the method to achieve BM reduction. We choose to express the problem as eigenvalue problems that will naturally extend to integro-eigenvalue problems in the continuous time case.
\begin{enumerate}
\item \label{step:itemeigenvalues}Solve the eigenvalue problem,
\beq
\label{eqs:BMAlgorithm1}
\begin{split} 
CC^\dagger U &= U C_D^2\\
C_{tt'}C^*_{t''t'}U_{t''n} &= U_{tm}(C_D)^2_{mn} \, .
\end{split}
\eeq
Solving this system gives the basis functions $\Psi_{nt} = U^*_{tn}$.
\item \label{step:rightsingularC} This unitary is composed of left singular vectors of C and leads to the corresponding unitary matrix of right singular vectors $V_C$, by,
\beq
\label{eqs:BMAlgorithm2}
\begin{split} 
C_D^{-1}U^\dagger C&=  V_C^\dagger \\
(C_D)_{nm} U^\dagger_{mt}C_{tt'} &= (V_C^\dagger)_{nt'}
\end{split}
\eeq
leading to the basis functions $\Phi_{nt} = (V_C)^*_{tn}$.
\item \label{step:Ssolutions}  We next find the right singular partial unitary $\tilde V_S$ corresponding to diagonalisation of $S$ using $\tilde U_C$, by first evaluating $\tilde S_D$ via $\tilde C_D^2-\mathds{1}=\tilde S_D^2$,where here the tilde indicates we take only the elements of $C_D$ which are greater than 1. We can then find the right partial unitary $\tilde V_S$ by,
\beq
\label{eqs:BMAlgorithm3}
\begin{split} 
\tilde S_D^{-1} \tilde U_C^\dagger S&=  \tilde V_S^\dagger \\
(\tilde S_D)_{nm} \tilde U^\dagger_{mt} S_{tt'} &= (\tilde V_S^\dagger)_{nt'}
\end{split}
\eeq

\item \label{step:Takagi} Finally, evaluate the Takagi decomposition of the matrix $\tilde G = \tilde V_C^\dagger \tilde V_S^*$ to give $\tilde G=\tilde D \tilde D^T$ and set,
\beq
\label{eqs:BMAlgorithm4}
\begin{split} 
\mathcal{U} &= U (\tilde D\oplus \mathds{1})\\
\mathcal{V}&= V_C (\tilde D\oplus \mathds{1})
\end{split}
\eeq
so that the Bloch-Messiah reduction is achieved with 
\beq
\label{eqs:BMAlgorithmResult}
\begin{split} 
C&= \mathcal{U}C_D\mathcal{V}^\dagger\\
S&= \mathcal{U} S_D \mathcal{V}^T  =  \mathcal{U} \tilde  S_D  \mathcal{V}^T
\end{split}
\eeq

\end{enumerate}

\section{Tensor Bloch-Messiah Reduction}
\label{app:GBMR}

\subsubsection{Spatial Mode Tensors}
Where we've previously considered just mode operators $\vec b_\omega$ with spectral degrees of freedom, we will extend these to include a spatial index $x$, so that $\vec b_{\omega x}$ denotes the annhilation operator for a photon with frequency $\omega$ in spatial mode $x$. 
A general Bogoliubov transformation of such modes takes the form
\beq
\label{eqs:MultiModeBogoliubov}
\begin{split} 
 \vec b_{\omega x} &= C_{\omega x \omega' x'} \vec a_{\omega' x'} + S_{\omega x \omega' x'} \vec a^\dagger_{\omega' x'} \,,
\end{split}
\eeq

For convenience we will not flatten the vector in the direction we append the creation operators onto the annihilation operators, instead we will attach an index $c$. So that $\vec B$, containing creation and annihilation operators, has elements $\vec B_{\omega xc}$ with $c=1$  the annihilation operator $\vec b_{\omega x}$ and $c=2$ the corresponding creation operator. The commutation relations can be written,
\beq
\label{eqs:MultiModeCommutation}
\begin{split} 
[\vec B_{\omega xc},\vec B_{\omega 'x'c'}] = \delta_{\omega \omega'} \delta_{xx'}\Omega_{cc'} 
\quad \text{with}\quad \Omega = 
\left( \begin{array}{cc}
0  & 1\\
-1 & 0 \end{array} \right)
\end{split}
\eeq
We use a collon to indicate tensor multiplication by summation over all left and right acting indices,
\beq
\label{eqs:TensorMultiplication}
\begin{split} 
( M : \vec A)_{\omega xc} &= M_{\omega xc\omega' x'c'} \vec A_{\omega' x'c'}\\
( M : M')_{\omega xc\eta yd}& = M_{\omega xc \omega 'x'c'}M'_{\omega 'x'c'\eta yd}\\
(C:\vec a)_{\omega x}& = C_{\omega x\omega 'x'} \vec a_{\omega 'x'}
\end{split}
\eeq

Writing the full Bogoliubov transformation using tensor multiplication we have,
\beq
\label{eqs:MultiModeBogoliubovTensor2}
\begin{split} 
\vec B = M : \vec A
\end{split}
\eeq
The symplectic structure reads $M:\mathds{1}^{(t)}\otimes\mathds{1}^{(s)}\otimes\Omega : M^T=\mathds{1}^{(t)} \otimes\mathds{1}^{(s)}\otimes\Omega$, where the transpose is understood to act on tensors by, $C_{\omega x \omega'x'}^T = C_{\omega 'x'\omega x}$ and  $M_{\omega xc\omega 'x'c'}^T = M_{\omega 'x'c'\omega xc}$, so explicitly we have,
\beq
\label{eqs:SymplecticTensor}
\begin{split} 
  \delta_{\omega \eta} \delta_{xy} \Omega_{cd}&= M_{\omega xc\omega 'x'c'} \delta_{\omega'\eta'} \delta_{x'y'} \Omega_{c'd'} M_{\eta yd\eta'y'd'} \\
&=  M_{\omega xc \omega'x'c'} \Omega_{c'd'}  M_{\eta yd\omega'x'd'} \, ,
\end{split}
\eeq
which leads to the analogous conditions 
\begin{align}
 C:C^\dagger-S:S^\dagger &= \mathds{1} \label{eqs:AllTConstraints1}\\
 C:S^T-S:C^T &= 0 \label{eqs:AllTConstraints2} \\
 C^\dagger :C-S^T: S^* &= \mathds{1} \label{eqs:AllTConstraints3} \\
 C^\dagger: S-S^T:C^* &= 0 \label{eqs:AllTConstraints4} \, .
\end{align}
We note that so far this is entirely equivalent to the previous definitions, just with the matrix elements folded into tensors across a relevant partition, ie. one would arrive back at the previous conditions by merely flattening any pairs of indices $t$ and $x$.

\subsubsection{Higher-order Singular Value Decomposition (HOSVD)}
Ordinary Bloch-Messiah reduction here would constitute a basis change using broadband functions $\Phi_{n,\omega x}$ which depend on both the temporal and spatial degrees of freedom. This is exactly what we want to avoid, instead we aim to find broadband mode functions on the spatial degree of freedom independently to those of the spectral degrees of freedom thus treating them independently. 

Rather than diagonalise the flattened tensor $CC^\dagger$, and consider this in terms of the singular value decomposition of the flattened $C$, we are going to find a HOSVD decomposition of $S$ of the form,
\beq
\label{eqs:TwoDOFCDecomp}
\begin{split} 
S_{\omega x \omega' x'} &= u^{(t)}_{\omega n} u^{(s)}_{x m} S^\perp_{n m n' m'} v^{(t)}_{\omega' n'} v^{(s)}_{x' m'} \,,
\end{split}
\eeq

where $U^{(t)}$, $U^{(x)}$, $V^{(t)}$ and $V^{(x)}$ are unitary matrices and the tensor $S^\perp$ is \emph{all orthogonal}, meaning,
\beq
\label{eqs:AllOrthogonal}
\begin{split} 
S^\perp_{txt'x'}S^{\perp *}_{txt'x''} = 0 \quad \forall \,x'\neq x'' \, ,
\end{split}
\eeq
and similarly for all indices of the tensor.

We may consider the constraint \ref{eqs:AllTConstraints1} using
\beq
\label{eqs:Constraint1HOSVD}
\begin{split} 
(S:S^\dagger)_{txt'x'} &= \mathcal{S}_{sys'y'} \mathcal{S}^*_{s'y's''y''} 
 U^{(t)}_{ts}U^{(x)}_{xy}
  U^{(t)*}_{s''t'}U^{(x)*}_{y''x'} \, ,
\end{split}
\eeq
by $(U^{(t)}\otimes U^{(x)})^\dagger : S:S^\dagger: (U^{(t)}\otimes U^{(x)}) = \mathcal{S}:\mathcal{S}^\dagger$, and considering $\mathcal{C}:\mathcal{C}^\dagger - \mathcal{S}:\mathcal{S}^\dagger = \mathds{1}$ however in general this fails to induce all orthogonality of the corresponding $C$ tensor. Consequently there is no equivalent step to the Takagi factorisation in the conventional method. We take as (possibly not all-orthogonal) $C^\perp$ the transformed kernel,
\beq
\label{eqs:TwoDOFCDecomp}
\begin{split} 
C^\perp_{n m n' m'} &= u^{(t)*}_{n \omega } u^{(s)*}_{ m x} C_{\omega x \omega' x'} v^{(t)*}_{n' \omega' } v^{(s)*}_{m' x'} \,.
\end{split}
\eeq

The practicality of such a basis is instead driven by the ordering of the elements of $S^\perp$ by their Frobenius norms which for certain figures of merit allow for optimal truncating of the system onto finite dimensional subspaces. In particular, the basis elements are ordered so as to maximize the total photon number (when acting this channel on the vacuum) in each mode.

\section{Antoine-Takiagi Reduction: Bloch-Messiah at The Hamiltonian Level}
\label{app:antoinetakagi}
In the case that the propagator takes on the form,
\beq
\label{eqs:PropagatorInteractionPic}
\begin{split} 
\hat{\mathds{U}} &= \exp\bigl\{-i \hat A^\dagger \cdot \bold{H} \cdot \hat A \bigr\}\\
&= \exp\bigl\{-i \left( \begin{array}{c}
\vec a\\
\vec a^{(\dagger)} \end{array} \right) ^\dagger \left( \begin{array}{cc}
0  & H\\
H^* & 0 \end{array} \right)\left( \begin{array}{c}
\vec a\\
\vec a^{(\dagger)} \end{array} \right)  \bigr\}
\end{split}
\eeq
(such as for interaction picture Hamiltonians neglecting time ordering), where $B$ can be chosen symmetric, $B=B^T$. A passive unitary transformation is achieved via,
\beq
\label{eqs:UnitaryOnPropagator}
\begin{split} 
\hat{\mathds{U}} &\Rightarrow \exp\bigl\{-i \hat A^\dagger \cdot
 \left( \begin{array}{cc}
U^\dagger  & 0\\
0 & U^T \end{array} \right)
\left( \begin{array}{cc}
0  & H\\
H^* & 0 \end{array} \right)
 \left( \begin{array}{cc}
U  & 0\\
0 & U^* \end{array} \right)
\bigr\} \cdot \hat A \\
&= \exp\bigl\{-i \hat A^\dagger \cdot
\left( \begin{array}{cc}
0  & U^\dagger H U^*\\
U^T H^* U & 0 \end{array} \right)
  \bigr\}\cdot \hat A\\
  &= \left( \begin{array}{cc}
U  & 0\\
0 & U^* \end{array} \right) \exp\bigl\{-i \hat A^\dagger \cdot
\left( \begin{array}{cc}
0  & H\\
H^* & 0 \end{array} \right)
  \bigr\} \left( \begin{array}{cc}
U^\dagger  & 0\\
0 & U^T \end{array} \right)\cdot \hat A\\
\end{split}
\eeq
The mode operators evolve as,
\beq
\label{eqs:SymplecticPropagator}
\begin{split} 
\hat{\mathds{U}} \hat A \hat{\mathds{U}}^\dagger &= \exp\bigl\{-I \bold K  \bold{H} \bigr\} \hat A\\
\end{split}
\eeq
with $\bold K$ the symmplectic form $\bold K = \text{diag}(1,1...,1,-1,-1,..,-1)$. If we take $U$ to be the unitary matrix diagonalising $H=UH^DU^T$, then we have,
\beq
\label{eqs:DiagPropagator}
\begin{split} 
 \exp\bigl\{-I \bold K  \bold{H} \bigr\} \hat A
&=\left( \begin{array}{cc}
U^\dagger  & 0\\
0 & U^T \end{array} \right)\left( \begin{array}{cc}
\cosh H^D  & -i \sinh H^D \\
i \sinh H^D & \cosh H^D \end{array} \right)  \left( \begin{array}{cc}
U  & 0\\
0 & U^* \end{array} \right) \hat A \, .
\end{split}
\eeq
We see therefore that Antoine-Takagi reduction leads to conventional Bloch-Messiah reduction of the linear symplectic transformation. 

\subsection{Generalised Antoine-Takiagi Reduction}
\label{app:generalisedantoinetakagi}
Performing HOSVD on the Hamiltonian level,
\beq
\label{eqs:AllOrth}
\begin{split} 
H_{txt'x'} =  U^{(t)}_{ts}U^{(x)}_{xy} H^\perp_{sys'y'}U^{(t)}_{s't'}U^{(x)}_{y'x'}
\end{split}
\eeq
leads in general to a reduction of the transformation which is inequivalent to GBM. Although the exponential map preserves the diagonal form of its argument, ie. Antoine-Takagi reduction simultaneously achieves Bloch-Messiah reduction, all-orthogonality of the Hamiltonian is not necessarily preserved. Consequently, a generalised Antoine-Takagi reduction is inequivalent to a generalised Bloch-Messiah reduction. In the case of two-mode squeezing however, the form of the Hamiltonian ensures these reduction coincide. Similarly, where the propagator is to be expanded only to first order, these reductions are again equivalent. The basis achieved by a generalised Antoine-Takagi reduction, is optimal for maximising the fidelity of a finite-dimensional approximation to the full biphoton state. 

\section{Single-mode Squeezing with loss}
Consider a Gaussian transformation on two spatial modes initially in the vacuum state, where spatial mode one, $x=1$, is a bus mode we have access to, and spatial mode two, $x=2$, is an inaccessible loss mode. A general transformation can be reduced to BM form~\ref{eqs:BMAlgorithmResult}resulting in basis functions with support across both spatial modes. To observe squeezing we transform to the quadrature basis $\hat X$ by $\hat X = \mathcal{Q} \hat A$, where
\beq
\label{eqs:ToQuad}
\begin{split} 
 \quad \mathcal{Q} &= Q^{(o)} \otimes \mathds{1}^{(ts)} \qquad \text{for} \qquad  Q^{(o)}= \left( \begin{array}{cc} 1&1\\-i&i \end{array} \right)\, ,
\end{split}
\eeq
with $ \mathds{1}^{(ts)}$ the identity on the temporal and spacial DOFs. The quadrature covariance matrix (given vacuum inputs) in the BM basis is given by $ {\tilde \sigma}_{nn'} = \mc M^D_{nm} \mc M^{D}_{n'm}$, with $\mathcal{M}^{D}_{nm} = (\mathcal{Q}M^{D}\mathcal{Q}^\dagger)_{nm}$.
Given access to both spatial modes, we would find the maximum squeezing in the minimum variance quadrature, $ {\tilde \sigma}_{11}$ . With access to only the bus mode we must expand the modal basis to respect the spatial structure, however, the impracticality of the BM basis lays in the fact that the partial basis modes $\{u_{n\omega 1}\}_n$ on spatial mode $x=1$ are not orthogonal, so we must Gram-Schmidt this set of modes to find the orthonormal set $\{ \bar u_{n\omega 1} = T_{n1n'1} u_{n' \omega 1}\}_n$ , and similarly for $\{u_{n \omega 2}\}_n$ to find $T_{n2n'2}$.
Under the full transformation our BM basis becomes $\bar B_{nx} =T \tilde B$.
The covariance matrix becomes ${\bar \sigma}_{nxn'x'} = (\mathcal{T} \mc M^D)_{nxmy}(\mathcal{T}\mc M^{D})_{n'x'my}$ where $\mathcal{T} = \mc Q T \mc Q^\dagger $.
Having access to only, $\bar \sigma^{(x=1)} = \{ {\bar \sigma}_{n1n'1}\}_{nn'} $, we observe maximum squeezing for some spectral mode $u^{(t)}$ on spatial mode $x=1$ by,\footnote{Whilst the quadrature basis symplectic transformation $\mathcal{U}^{(t)}$ must be of the general form $\mathcal{U}^{(t)} = \arFour{\Re(U^{(t)})}{-\Im(U^{(t)})}{\Im(U^{(t)})}{\Re(U^{(t)})}$ (with $\Re(\cdot)$ ($\Im(\cdot)$) the Real (Imaginary) part), for some unitary $U^{(t)}$, it is always possible to efficiently determine the minimum eigenvalue and corresponding eigenvector of $\bar \sigma$ then complete $\mathcal{U}^{(t)}$ to a valid orthogonal symplectic matrix.}
\beq
\label{eqs:SingleModeminvar}
\begin{split} 
\argmin_{u^{(t)}} \bra{u^{(t)}} \bar \sigma^{(x=1)}\ket{ u^{(t)}} \,.
\end{split}
\eeq

In contrast, we now consider the GBM form applied to this system. The core tensor in the quadrature basis is, $\mathcal{M}^{\perp}_{nxn'x'} = (\mathcal{Q}M^{\perp}\mathcal{Q}^\dagger)_{nxn'x'}$, (where $\mc Q = Q^{(o)}\otimes \mathds{1}^{(t)} \otimes \mathds{1}^{(s)}$) giving the covariance tensor $\breve \sigma_{nxn'x'} = \mc M^\perp_{nxmy}\mc M^\perp_{n'x'my}$, and by applying  $\mc U^{(s)\,-1}$ we arrive at $\breve \sigma^{(x=1)} = \{ ( \mc U^{(s)\dagger}\breve \sigma  \mc U^{(s)})_{n1n'1}\}_{nn'} $, to which we may perform the spectral mode minimisation as per Eqs.~\ref{eqs:SingleModeminvar}.

\section{Equivalence of Two-mode and Single-mode Squeezing Pictures}
\label{app:twomodeequivalence}
Two independent single mode squeezers,
\beq
\label{eqs:SinglemodeSqueeze}
\begin{split} 
\vFour{\hat B_1}{\hat B_2}{\hat B_1^\dagger}{\hat B_2^\dagger}
= \left( \begin{array}{cccc} {c} & {0} & {s} & {0} \\ 
{0} & {c} & {0} & {s} \\
{s^*} & {0} & {c^*} & {0} \\
{0} & {s^*} & {0} & {c^*}  \end{array} \right) \vFour{\hat A_1}{\hat A_2}{\hat A_1^\dagger}{\hat A_2^\dagger} \, ,
\end{split}
\eeq
and a two mode squeezer,
\beq
\label{eqs:TwomodeSqueeze}
\begin{split} 
\vFour{\hat {B'}_1}{\hat {B'}_2}{\hat {B'}_1^\dagger}{\hat {B'}_2^\dagger}
= \left( \begin{array}{cccc} {c} & {0} & {0} & {s} \\ 
{0} & {c} & {s} & {0} \\
{0} & {s^*} & {c^*} & {0} \\
{s^*} & {0} & {0} & {c^*}  \end{array} \right) \vFour{\hat {A'}_1}{\hat {A'}_2}{\hat {A'}_1^\dagger}{\hat {A'}_2^\dagger} \, .
\end{split}
\eeq
%
differ by a passive unitary, $U$, equivalent to a balanced beamsplitter, such that $UU^T=X$, with $X$ the Pauli X matrix, (which can be found by Takagi factorisation), since
\beq
\label{eqs:TwomodetooSinglemode}
\begin{split} 
 \left( \begin{array}{cccc} {c} & {0} & {0} & {s} \\ 
{0} & {c} & {s} & {0} \\
{0} & {s^*} & {c^*} & {0} \\
{s^*} & {0} & {0} & {c^*}  \end{array} \right)  = \arFour{U}{0}{0}{U^*} \left( \begin{array}{cccc} {c} & {0} & {s} & {0} \\ 
{0} & {c} & {0} & {s} \\
{s^*} & {0} & {c^*} & {0} \\
{0} & {s^*} & {0} & {c^*}  \end{array} \right)\arFour{U}{0}{0}{U^*}^\dagger
\end{split}
\eeq

We next consider a general case arising from a block-anti-diagonal squeezer,
%
\beq
\label{eqs:STwoModeGeneralCombined}
\begin{split} 
C& = \arFour{Q^{(ss)}}{0}{0}{Q^{(ii)}}\\
 S&= \arFour{0}{Q^{(si)}}{Q^{(is)}}{0}
\end{split}
\eeq
Singular value decomposition of $C$ can be achieved by block diagonal unitaries,
\beq
\label{eqs:CTwoModeGeneralCombinedSVD}
\begin{split} 
C& = \arFour{U_1}{0}{0}{U_2}\arFour{Q^{D(ss)}}{0}{0}{Q^{D(ii)}}\arFour{V_1}{0}{0}{V_2}^\dagger \, ,
\end{split}
\eeq
whilst the symplecticity conditions (Eqs.~\ref{eqs:AllConstraints1}~-~\ref{eqs:AllConstraints4} ) ensure that the terms in $S$ are diagonalised by,
\beq
\label{eqs:STwoModeGeneralCombinedSVD}
\begin{split} 
S& = \arFour{U_1}{0}{0}{U_2}\arFour{0}{Q^{D(si)}}{Q^{D(is)}}{0}\arFour{V_1}{0}{0}{V_2}^T \, ,
\end{split}
\eeq
and that the singular value spectrum of, $Q^{D(ss)}$ and $Q^{D(ii)}$, (equivalently $Q^{D(si)}$ and $Q^{D(is)}$) are equal.
This results in decoupled equations for each independant two mode squeezer upon which we can apply the above transformation Eqs.~\ref{eqs:TwomodetooSinglemode} (with $U$ expanded over the multiple spectral modes $U\otimes \mathds{1}$).

Furthermore, in the case of an Antoine-Takagi decomposition of a Hamiltonian of the form,
\beq
\label{eqs:TwoModeHamiltonian}
\begin{split} 
H=\arFour{0}{F_{JSA}}{F^T_{JSA}}{0} \, 
\end{split}
\eeq
$F_{JSA}=U_1 F_{JSA}^D U_2^T$ means the transformation is symmetric $U_1=V_1$, $U_2=V_2$, and the squeezing parameters relate to the Hamiltonian via $Q^{D(ss)}=Q^{D(ii)}= \cosh (F_{JSA}^D)$ and $Q^{D(si)}=Q^{D(is)}= \sinh (F_{JSA}^D)$

\section{Truncating Transformations}
We may without loss of generality, first perform a polar decomposition to an arbitrary transformation, $M = M' \mathds{V}$, consider the truncations of the symmetric active component, $M'$, then reintroduce the orthonormal transformation, $\mathds{V}$, afterwards. Consider then a general symmetric transformation in the GBM basis,
\beq
\label{eqs:SymmetricMGBM}
\begin{split} 
M' =  \left( \begin{array}{cccc} {\tilde C} & {C_{12}} & {\tilde S} & {S_{12}} \\ 
{C_{12}^T} & {\bar C} & {S_{12}^T} & {\bar S} \\
  {\tilde S^*} & {S_{12}^*} &{\tilde C^*} & {C_{12}^*} \\ 
 {S_{12}^\dagger} & {\bar S^*} &{C_{12}^\dagger} & {\bar C^*} \end{array} \right)  
\end{split}
\eeq
Consider the total number operator on the $d$ dimensional subspace on which $\tilde S$ acts, $\hat N^{(d)} = \sum_i^d \hat n_i$. When $M'$ acts on the vacuum we have
\beq
\label{eqs:NdExpectation}
\begin{split} 
\av{\hat N^{(d)}}_{M'} &= \sum_{i}^d \sum_j |S_{ij}|^2\\
&= |\tilde S|_F^2 + |S_{12}|_F^2
\end{split}
\eeq
If $M'$ is in the GBM basis, then this expectation value is maximised by definition. 


\section{Review of Methods for HOSVD}
\label{app:HOSVD}
We review how one achieves HOSVD by SVD of flattened tensors~\cite{DeLathauwer2000}. 

To begin, we define the $\bold n$-mode flattening of a tensor $A$. Take $A\in \mathbb{C}^{I_1\times I_2\times...\times I_N}$, then $A_{(\bold n)}$ is found by flattening the tensor in all directions except $\bold n$ and is thus the matrix,
\beq
\label{eqs:nmodeflattening}
\begin{split} 
A_{(\bold n)} \in \mathbb{C}^{I_{n} \times I_{n+1}I_{n+2}...I_{N}I_{1}...I_{n-1}}
\end{split}
\eeq
with element $A_{i_1i_2,...i_N}$ at row number $i_{\bold n}$ and collumn number 
\beq
\label{eqs:nmodeflatteningcollumn}
\begin{split} 
(i_{n+1}-1)I_{n+2}...I_NI_1I_2,...I_{n-1} + (i_{n+2}-1)I_{n+3}...I_NI_1I_2,...I_{n-1} +\cdots +i_{n-1}
\end{split}
\eeq

These unfolding allow us to define the unitaries matrices from the HOSVD as terms in the SVD of matrix unfoldings through,
\beq
\label{eqs:nmodeSVD}
\begin{split} 
A_{(\bold n)} =U^{(\bold n)}  \Sigma^{(\bold n)} V^{\bold n \dagger}
\end{split}
\eeq
and the all-orthogonal core tensor, $\mathcal{A^\perp}$, can be recovered via,
\beq
\label{eqs:nmodeSVDCoreTensor}
\begin{split} 
\mathcal{A^\perp} = A :   U^{\dagger (1)}\times U^{\dagger (2)}\times \cdots \times U^{\dagger (N)}
\end{split}
\eeq
Resulting in the HOSVD of $A$,
\beq
\label{eqs:nmodeSVDCoreTensor}
\begin{split} 
A = \mathcal{A^\perp} :   U^{ (1)}\times U^{ (2)}\times \cdots \times U^{ (N)}\, .
\end{split}
\eeq
\end{appendices}

\addcontentsline{toc}{chapter}{Bibliography}

\bibliography{library}
\bibliographystyle{apsrev4-1}

\end{document}